# Hydration induced spin glass state in a frustrated Na-Mn-O triangular lattice

Ioanna Bakaimi,[1,2] Rosaria Brescia,[3] Craig M. Brown,[4,5] Alexander A. Tsirlin,[6] Mark A. Green,[7] and Alexandros Lappas[1,*]

[1] *Institute of Electronic Structure and Laser, Foundation for Research and Technology – Hellas, Vassilika Vouton, 71110 Heraklion, Greece*

[2] *Department of Physics, University of Crete, Voutes, 71003 Heraklion, Greece*

[3] *Nanochemistry Department, Istituto Italiano di Tecnologia, Via Morego 30, 16163 Genova, Italy*

[4] *NIST Center for Neutron Research, 100 Bureau Drive, Gaithersburg, MD 20899-8562, USA*

[5] *Department of Chemical and Biomolecular Engineering, University of Delaware, Newark, DE 19716, USA*

[6] *Experimental Physics VI, Center for Electronic Correlations and Magnetism, Institute of Physics, University of Augsburg, 86135 Augsburg, Germany*

[7] *School of Physical Sciences, University of Kent Canterbury, Kent CT2 7NH, UK*

## Abstract

Birnessite compounds are stable across a wide range of compositions that produces a remarkable diversity in their physical, electrochemical and functional properties. These are hydrated analogues of the magnetically frustrated, mixed-valent manganese oxide structures, with general formula, $Na_xMnO_2$. Here we demonstrate that the direct hydration of layered rock-salt type α-$NaMnO_2$, with the geometrically frustrated triangular lattice topology, yields the birnessite type oxide, $Na_{0.36}MnO_2·0.2H_2O$, transforming its magnetic properties. This compound has a much-expanded interlayer spacing compared to its parent α-$NaMnO_2$ compound. We show that while the parent α-$NaMnO_2$ possesses a Néel temperature of 45 K as a result of broken symmetry in the $Mn^{3+}$ sub-lattice, the hydrated derivative undergoes collective spin-freezing at 29 K within the $Mn^{3+}$/$Mn^{4+}$ sub-lattice. Scaling-law analysis of the frequency dispersion of the AC susceptibility, as well as the temperature-dependent, low-field DC magnetization confirm a cooperative spin-glass state of strongly interacting spins. This is supported by complementary spectroscopic analysis (HAADF-STEM, EDS, EELS) as well as by a structural investigation (high-resolution TEM, X-ray and neutron powder diffraction) that yield insights into the chemical and atomic structure modifications. We conclude that the spin-glass state in birnessite is driven by the spin-frustration imposed by the underlying triangular lattice topology that is further enhanced by the in-plane bond-disorder generated by the mixed-valent character of manganese in the layers.

* Electronic address: lappas@iesl.forth.gr

**I. INTRODUCTION**

Layered manganese oxides containing triangular magnetic lattice topology have demonstrated important phenomena including magnetic frustration [1], unique electronic properties as a result of the Jahn-Teller active $Mn^{3+}$ cations [2], and the development of different polymorphic phases [3] with variable magnetic response and redox potentials, overall reflecting a notable interplay between microstructure and properties. Due to their modular porosities, high thermal surface area and stabilities, these compounds find numerous technological applications such as their use in ion sieves, molecular sieves, catalysts [4,5] and electrodes in batteries. One of the most prominent characteristics of manganese oxides is that their structure can be used as a host material for intercalation and de-intercalation reactions. $LiMnO_2$ [6] is a member of the manganese oxides family, which is widely known for its use as cathode material in Li rechargeable batteries. The replacement of the conventional Li- based compounds with Na- based ones, such as birnessite, is of great importance due to the higher abundance and relative low cost of sodium compared with lithium. Examples of sodium manganese oxides which have been studied as candidates for the aforementioned applications are the $Na_{0.44}MnO_2$ [6,7] and recently reported β-$NaMnO_2$ [8] where tuning the concentration of planar defects [3] determines its quality as a high performance cathode material for battery technologies[9].

A common characteristic of layered manganese oxides is the presence of edge shared $MnO_6$ octahedral units, which are separated by metal ions such as Li, Na or K [10]. The interlayer distance varies between particular structures from ~4.7 Å in $Li_{1.09}Mn_{0.91}O_2$ [11], to ~7 Å in the birnessite [12] and ~10 Å in the buserite structure [13]. The hydrated $NaMnO_2$ oxides, whose Na/Mn molar ratio is between 0.2-0.7, are known as Na-birnessites [10,14]. Depending on the intercalating cation there are also K, Li [15], Bi, or Zn based birnessites. The latter have recently been discussed to have a critical role in Zn-batteries during the charging process [16]. Interestingly, a recent study showed that hydrated manganese oxides such as the $Na_{0.71}MnO_2 \cdot 0.25H_2O$, have superior cycling stability and high capacity as a result of the presence of water within the interlayer gap [17].

There are several studies [11,12,15,18-20] describing the preparation methods of birnessites including deintercalation [19], ion exchange [19] and hydrothermal synthesis [18]. Here, we evaluate Na-birnessite prepared from direct hydration of pure geometrically frustrated α-NaMnO$_2$. The latter is a rock salt type derivative (C2/m, a= 5.67 Å, b= 2.85 Å, c= 5.80 Å, β=113.2°) which may be used as a precursor for the preparation of the cathode material LiMnO$_2$. Its structure is composed of sheets of edge-sharing MnO$_6$ octahedra, which are separated by a single two-dimensional layer of Na cations. Geometric magnetic frustration develops as the Jahn-Teller (J-T) active $Mn^{3+}$ ($3d^4$, S=2) distorts the in-plane Mn-topology allowing for a spatially anisotropic $J_1$-$J_2$ two-dimensional (2D) triangular spin lattice [1]. When spin frustration is lifted below 45 K, this system adopts a remarkable nanoscale inhomogeneity in the ground state, with Néel order being the outcome of local symmetry-breaking pinning sites [21]. A recent study highlights that the formation of different NaMnO$_2$ polymorphs, such as α-NaMnO$_2$ and β-NaMnO$_2,$ is largely driven by the evolution of relative concentration of planar defects in their microstructure [3]. Moreover, alkali substoichiometry, as in α-Na$_{0.7}$MnO$_{2.25}$ [22], allows mixing $Mn^{3+}$ and $Mn^{4+}$ states, therefore determining the important role of structural details in modifying the cooperative nature of the electronic/magnetic properties. The latter is demonstrated in a spectacular manner with the interplay of Na-vacancies and cooperative J-T effects in Na$_{5/8}$MnO$_2$ [23],where a favorable charge ordered state is established. However, despite the plethora of the synthesis-based studies found in the literature, there are only few reports related to the magnetic properties of the birnessite-based systems [24]. One such study is the reported work on the birnessite type MnO$_2$ nanowalls [24] that were found to exhibit a magnetic transition at 9.2 K, evident by a bifurcation of the DC susceptibility zero field cooled and field cooled (ZFC-FC) curves. The magnetic properties of the birnessite type nanowalls were attributed to the antiferromagnetic interactions between the $Mn^{3+}$/$Mn^{4+}$ cations [24].

In this work, we examine the impact of the structure and composition modifications on the magnetism after intercalating water-molecules between the $MnO_6$ layers of α-$NaMnO_2$. The hydration of the latter resulted in the formation of birnessite-type material with the composition $Na_{0.36}MnO_2·0.2H_2O$, and concomitant enhancement of the interlayer spacing. This drastically changes the magnetic properties and the antiferromagnetic ordering of the parent α-$NaMnO_2$ gives way to a strongly interacting spin glass state. Indeed, extensive DC and AC magnetic susceptibility studies corroborate that the system undergoes spin glass freezing below 29 K. We demonstrate that the frequency dispersion of the temperature-dependent AC susceptibility maximum is described well by the dynamic scaling analysis according to critical slowing down (a phenomenological power-law behavior). Based on the experimental evidence we discuss that frustration caused by bond disorder, attributed to competing exchange interactions due to random distribution of the $Mn^{3+}/Mn^{4+}$ cations, provides the conditions for such an emerging state.

**II. EXPERIMENTAL PROCEDURES**

Polycrystalline powder of $Na_{0.36}MnO_2·0.2H_2O$ (hereafter referred as Na-birnessite in abbreviation) was synthesized by direct hydration of α-$NaMnO_2$[1] and based on a synthesis protocol previously reported [25]. The starting material α-$NaMnO_2$ was prepared by solid state reaction: stoichiometric amounts of $Na_2CO_3$ and $Mn_2O_3$ were mixed, ground, pelletized and heated from room temperature up to 750 °C, with a heating rate of 3 °C/min under Argon atmosphere [3]. After holding at high temperature for 60 hrs the pellet was cooled to room temperature. The final product was exposed to ambient atmosphere for two weeks. This ensured the homogeneous hydration of α-$NaMnO_2$ within the interlayer gaps, which leads to the formation of the Na-birnessite compound. Frequent mixing of the powder was undertaken so that all crystallites would become exposed to air. This promoted the homogeneity of the final compound, while the powder converted from the host framework α-$NaMnO_2$ to the Na-birnessite.

X-Ray powder diffraction (XRPD) experiments were carried out on a Rigaku [26] D/MAX-2000H rotating Cu anode diffractometer (λ=1.5406 Å). Samples were prepared for TEM investigations by mild sonication in ethanol and drop-casting onto Cu grids covered with a holey carbon film. High-resolution transmission electron microscopy (HRTEM), electron energy-loss spectroscopy (EELS) and energy-dispersive X-ray spectroscopy (EDS) analyses were carried out using a Jeol JEM 2200FS instrument, equipped with a Schottky emitter operated at 200 kV, a CEOS spherical aberration corrector of the objective lens allowing for a spatial resolution of 0.9 Å, and an in-column imaging filter (Ω-type). EELS analyses were carried out in TEM mode (convergence and collection angles 5.5 mrad and 1 mrad, respectively, 0.4 eV/pixel dispersion) and the quantification was carried out using EELS Model [27]. EDS mapping and compositional quantification was determined in scanning TEM (STEM)-high angle annular dark field (HAADF) imaging mode, using a Bruker Quantax 400 system with a 60 $mm^2$ XFlash 5060 silicon drift detector. Thermal analysis (Thermogravimetric – TGA and Differential Thermal Analysis – DTA) was performed on a SDT-Q600 TA instruments system under Ar-gas flow. Neutron Powder Diffraction (NPD) data were collected using the high-resolution powder diffractometer BT-1 at the National Institute of Standards and Technology Center for Neutron Research (NCNR), with a wavelength of 1.5406 Å and 60' collimation from the Cu-311 monochromator.

Magnetic susceptibility (DC and AC) was measured on a Superconducting Quantum Interference Device (SQUID) magnetometer (Quantum Design MPMS-XL5) under various protocols and magnetic fields (H= 2.5 mT – 100 mT). The frequency dependent (f= 47 Hz – 901 Hz) AC susceptibility data, which were analyzed by appropriate phenomenological laws, were collected using the temperature sweep mode of the MPMS. This mode was chosen as a more appropriate one, since it affects less the thermodynamic and time dependent phenomenon of the encountered spin glass transition.

## III. RESULTS AND DISCUSSION

### A. Transmission Electron Microscopy studies (EDS, HRTEM, EELS)

EDS quantification has been performed in STEM mode by the Cliff-Lorimer method, combined with HAADF-STEM imaging to determine the distribution of the Na, Mn and O atoms in the Na-birnessite sample. While EDS mapping typically shows a homogeneous distribution of the Mn, O and Na over individual crystals (Figure 1), atomic quantification over several crystals results in an average Na/Mn ratio of 0.36(±0.10). Due to the low accuracy of EDS for quantification of light elements, the O/Mn atomic ratio was instead obtained as 2.2(±0.5) from EELS analyses over three different crystals of the Na-birnessite. According to elemental analyses, the chemical formula corresponding to the hydrated compound was then estimated to be $Na_{0.36}MnO_2 \cdot 0.2H_2O$.

Together with the aforementioned results of compositional analysis, a HRTEM study of individual crystallites was fundamental for the identification of the crystal structure of the Na-birnessite phase. The images reveal the presence of μm scale crystals in the hydrated compound whose structure matched well that reported for the $Na_{0.3}MnO_2 \cdot 0.93H_2O$ [28] birnessite. Figure 2 (and Figure S1) presents one such crystal which has been used for the analysis. The fast Fourier transform (FFT; Fig. 2b) of the area in Fig. 2c matches to the [00-1] orientation of the $Na_{0.3}MnO_2 \cdot 0.93H_2O$ phase, with a 6% dilation compared to the reported cell parameters. According to the FFT, analysis the hydrated compound crystallizes in the triclinic system (space group $C\bar{1}$), with the following cell parameters: a= 5.53(1) Å, b= 3.11(6) Å, c= 7.80(1) Å, α= 89.492(13)°, β= 103.136(12)°, γ= 89.929(10)°. Stacking faults appear parallel to the {210} planes (Fig. 2c). These structural defects are not parallel to the {100} planes, in which $H_2O$ molecules are inserted in the structure. The structural modifications could result from Mn vacancies. Noticeably, the c parameter has increased in the hydrated final product by ~2 Å in comparison with that of the parent host α-$NaMnO_2$ (c~ 5.8 Å, at room temperature). This is nicely illustrated by means of XRPD (Figure 3), in which the 001 Bragg reflection of α-$NaMnO_2$ is shifted towards lower angles when the Na-birnessite compound is formed. A schematic representation of the ideal structure shown in Figure 4 depicts the hydrated material's in-plane triangular Mn sub-lattice topology.

Furthermore, the manganese $L_{2,3}$ core-loss EEL spectrum of the Na-birnessite was recorded (Figure S2) and carefully analyzed based on the white line ratio method [29]. According to this method, the integral intensity ratio of the $L_3$ and $L_2$ excitation peaks of a transition metal is correlated to its formal oxidation state [29-35]. The analysis resulted in the average oxidation state of 3.4, postulating a mixed valent ($Mn^{3+}/Mn^{4+}$) character for the manganese cations.

## B. Thermogravimetric Analysis

Measurements of the weight-loss versus temperature were carried out from room temperature up to 1100 °C with a small quantity (<10 mg) of the Na-birnessite sample that was heated up with a constant rate of 20 °C/min. The results are shown in Figure 5. Above 100 °C there is a 13% decrease in the weight up to 500 °C, which is attributed to the removal of $H_2O$ from the Mn-O sheets. Correspondingly the DTA graph exhibits a sharp endothermic peak around 140 °C due to the dehydration of the Na-birnessite. Further heating between 500 - 800 °C results in about 4 % reduction of the compound's weight that is accompanied by a broad dip in the DTA curve, which starts above 400 °C and an endothermic peak at 642 °C. The changes in this temperature regime may be caused by a transformation of birnessite to other layered polymorphs such as γ-$MnO_2$, and its subsequent partial conversion to $Mn_2O_3$, inferring reduction of the tetravalent manganese correlated also with the release of oxygen [18]. Due to the disproportionation of manganese valence state, analogous observation have been made in the TG-DTA measurements for Na-deficient α-$Na_{0.7}MnO_{2.25}$ [22], as well as other birnessite-like systems including ($H_{0.22}MnO_2·0.62H_2O$) [36], and $Na_4Mn_{14}O_{27}$ $9H_2O$ [14]. The additional weight loss (~2%) up to 1000 °C, likely relates to redox/extraction reactions due to the increased mobility of alkali ions in forming high-temperature manganese polymorphs. In view of this, powder XRD was performed on the product formed immediately after the TGA-DTA experiment, and compared to the as-synthesized compound. The XRPD patterns presented in Figure 6 indicate that the sample heated up to 1100 °C in the TGA reverts back to layered α-$NaMnO_2$, although impurities of β-$NaMnO_2$ and $Mn_3O_4$ phases are also present. The absence of the (001) reflection in Na-birnessite after the heating stage, confirms the elimination of intercalated $H_2O$ in the final product.

## C. Neutron Powder Diffraction

NPD patterns were taken at different temperatures, namely at 5, 20, 50 and 300 K (Figure 7), with the purpose to identify any possible phase transitions of either structural or magnetic origin. Although the high background due to the large incoherent scattering cross section of water protons may hinder low-angle diffuse scattering that would point to short-range or low-dimensional spin-correlations, it is important to note the lack of any additional magnetic Bragg scattering in the low-temperature patterns at 5 K and 20 K. This qualitative observation substantiates the absence of long-range magnetic order confirming that Na-birnessite features a magnetically disordered ground state, presumably stemming from spin-frustration caused by competing exchange interactions due to topology and/or site disorder. This is in contrast to the low-angle magnetic Bragg reflections revealed in Na-deficient $Na_xMnO_2$ (x= 5/8), where charge-ordering of Na and Mn stripes yield a fascinating low-temperature magnetic ordering below about 60 K [23].

**D. Static Magnetic Susceptibility**

The Na-birnessite is synthesized through chemical transformation of the α-$NaMnO_2$ host framework. Our studies provide new evidence for the substantial changes in the magnetic properties of α-$NaMnO_2$ when converted to Na-birnessite. Figure 8 compares the ZFC DC susceptibility (20 mT) for the two materials. The magnetic susceptibility of α-$NaMnO_2$ does not show a distinct magnetic transition, and rather a broad hump at high temperatures indicative of low dimensional magnetism is seen [37]. On the contrary, the magnetic susceptibility of Na-birnessite shows a sharp peak at 29 K, strongly suggesting that the compound undergoes a magnetic transition. We note that in the $AMnO_2$ rock-salt type related derivatives, solid-state NMR has shown that the orbital overlap and charge transfer from $Mn^{3+}$ to the interlayer cations (A= Cu, Na) is much larger in $CuMnO_2$ than in $NaMnO_2$ [38]. This corroborates to the role of interlayer species controlling the Néel state, setting in at 45 K for Na and at 65 K for Cu, however, accompanied by a differing ferromagnetic (FM) and antiferromagnetic (AFM) modification of the Mn-Mn interplane couplings, respectively. In view of these facts, the qualitatively different features in χ(T) and the nature of the magnetism in α-$NaMnO_2$ and Na-birnessite may be related to the critical modification of the interlayer Mn-Mn interaction strength along the c-axis, stemming from the different nature of the interlayer motifs accommodated therein. Furthermore, the static (DC) magnetic susceptibility of the Na-birnessite has been measured on the basis of a ZFC-FC protocol and under 20 mT external magnetic field (Figure 9). Below 29 K there is a significant bifurcation between ZFC and FC curves that seems to infer a collective spin freezing, and no long-range magnetic ordered state, in accord with the absence of magnetic Bragg peaks verified by NPD below this transition point. It is worth noting that many compounds which undergo a spin-glass transition, such as the $La_{0.5}Sr_{0.5}CoO_3$ [39], $CuFe_{0.5}V_{0.5}O_2$ [40], $Co_{1-x}Mn_xCl_2·H_2O$ [41], and most importantly, the archetypal spin-glass CuMn [42] lack long-range magnetic order, but exhibit both the sharp maximum in the ZFC susceptibility, $\chi_{ZFC}(T)$, as well as the divergence between the ZFC and FC curves below the $T_f$ (where $T_f$ stands for the spin-freezing temperature). In the following (sections E and F) we will provide more detailed evidence for the spin-glass magnetism in Na-birnessite.

Fitting of $\chi(T)$ for Na-birnessite on the basis of the Curie-Weiss law $\chi = C/(T - \theta_{cw})$ is presented in the inset of Figure 9. The derived values provide an estimate for the paramagnetic effective moment, $\mu_{eff}$ = 3.401(2) $\mu_B$ and the Weiss temperature, $\theta_{cw}$ = -64.28 (1) K. It is intriguing to analyze how the $\mu_{eff}$ for the Na-birnessite is justified in view of the formal spin-only effective moments of $Mn^{3+}$ ($d^4$) in its low-spin ($t_{2g}^4 e_g^0$; S= 1; 2.83 $\mu_B$) and high-spin ($t_{2g}^3 e_g^1$; S= 2; 4.9 $\mu_B$) configurations, as well as that of $Mn^{4+}$ ($t_{2g}^3 e_g^0$; S= 3/2; 3.87 $\mu_B$). For example, in the chemically related $Na_{0.7}MnO_2$ the effective moment was estimated at 4.77 $\mu_B$ ($\theta_{cw}$ = -411 K), closely resembling $Mn^{3+}$ in a high spin-state [43]. The $\mu_{eff}$ for the Na-birnessite, on the other hand, may be attributed to a combination of two available valence states for the manganese cations, whose ratio $Mn^{4+}/Mn^{3+}$ has been estimated as ~0.98 (Supplemental Material, S2) [44]. Literature reports suggest that when the $Mn^{4+}/Mn^{3+}$ ratio is above a critical value of 0.4-0.5 [45], $Mn^{3+}$ ions prefer the low-spin state; a case that appears to hold also for Na-birnessite. The unusual low-spin $Mn^{3+}$ state has also been predicted by DFT, and confirmed by magnetic susceptibility studies on Cr-doped rhombohedral $LiMnO_2$ ($\mu_{eff}$ = 2.97 $\mu_B$) [46]. Moreover, its occurrence in closely related porous manganese oxide octahedral sieves [47] has been claimed to emerge from the suppression of the Jahn-Teller distortion caused by the influence of higher valence cations such as $Mn^{4+}$. It is worth noting that in the parent oxide α-$NaMnO_2$, which is not a mixed valent compound, the Jahn-Teller active $Mn^{3+}$ cations adopt the high-spin state alone [1], with antiferromagnetic long-range order at low temperatures. In view of the above, Na-birnessite and its chemically related dioxides, $Na_{0.7}MnO_2$ (hexagonal) [43] and α-$K_{0.087}MnO_2$ (tetragonal) [48], seem to support both site-disorder ($Mn^{3+}$ and $Mn^{4+}$ mixture) induced by sub-stoichiometry in the interlayer sites as well as frustration due to the underlying triangular sublattice that have both been claimed as the microscopic features central to the emerging spin glass-like phase transitions.

**E. Dynamic Magnetic Susceptibility**

The transition to the spin glass state is a dynamical process. Taking this into account, the signature for glassiness is derived and affirmed by AC magnetic susceptibility experiments. Figure 10 shows the real $\chi'(T)$ and imaginary $\chi''(T)$ parts of the susceptibility measured at five frequencies (f = 47, 77, 97, 217, 901 Hz) upon a ZFC mode under a 0.3 mT AC driving field. There are two notable characteristics. First, the peak observed in both the real and the imaginary parts of the magnetic susceptibility decreases with increasing frequency. Second, the frequency dependent transition temperature, $T_f$, shifts to higher values as the frequency increases, similarly with the magnetic dynamics of other known spin-glass systems, such as for example, the $Ge_{1-x}Mn_x$ [49] or the $La(Fe_{1-x}Mn_x)_{11.4}Si_{1.6}$ compounds [50]. In view of this, quantifying the frequency shift of the $T_f$, measured as the relative variation of $\chi'(T)$ peak-temperature position per frequency ($\omega = 2\pi f$) decade

$$K = \frac{\Delta T_f}{T_f \Delta(log\omega)} \qquad (1)$$

provides some phenomenological description on the strength of interactions and offers a good criterion for distinguishing a spin-glass behaviour from that of a superparamagnet. This is known as the Mydosh parameter [42,51], which for Na-birnessite yields K~ 0.007. It is worth noting that K falls in the range expected for spin-glasses (0.005< K< 0.06) [42,52-54] and is comparable with those values obtained for the prototypical, cooperative spin-glasses of CuMn (0.005) [42] and AgMn (0.006) [42]. On the contrary, K for superparamagnets is an order of magnitude larger, varying from 0.1-0.3 upon gradual blocking of moments [51]; for example K= 0.28 for the $\alpha$-$(Ho_2O_3)(B_2O_3)$ superparamagnet [42]. Therefore, the estimated Mydosh parameter excludes the scenario of the superparamagnetism for Na-birnessite.

It is well-accepted though to verify the spin-fluctuations and their strength against the frequency dispersion of the $\chi'(T)$ maximum. The description of the latter by the available phenomenological laws (vide infra) evaluates the temperature dependence of the spin relaxation time $\tau$, which in turn, provides information on the underlined dynamics. In view of these, the Arrhenius law is utilized to describe weakly or non-interacting magnetic moments [42]:

$$\tau = \tau_o \exp\left(\frac{E_a}{k_B T}\right) \quad (2)$$

where $E_a$ is the activation energy determined by the energy needed to exit out of a local potential well [55], $\tau_o$ is the "attempt" time [55],[56] and T stands for the frequency dependent $\chi'(T)$ peak-temperature position, $T_m$ from here onwards. However, the analysis on the basis of the Arrhenius law (Supplemental Material, S3) [44] resulted in physically unreasonable parameters, namely, $\tau_o \sim 10^{-151}$ sec ($\pm 10^{-159}$ sec) and $\frac{E_a}{k_B}$ = 10254(567) K (Supplemental Material, Figure S3) [44]. This failure adds further support to the argument that the spin dynamics in Na-birnessite is not dictated by superparamagnetic blocking of spins, but instead magnetic moments with elevated strength of interactions may come into play. For this, the Vogel-Fulcher law for intermediate strength interactions may be another appropriate candidate phenomenological description of spin-fluctuations [57]:

$$\tau = \tau_o \exp\left(\frac{E_a/k_B}{T_m - T_f}\right) \quad (3)$$

where $T_m$ is the frequency-dependent peak temperature for the real part of $\chi'(T)$, and $T_f$ is the spin-glass freezing temperature that corresponds to a qualitative estimate of the inter-site interaction energy strength. However, the best fit from the Vogel-Fulcher analysis resulted in $T_f$ = 29.64 K, $\frac{E_a}{k_B}$ = 2.09(1) K and an attempt time, $\tau_o$ = 5.09×10$^{-6}$ ($\pm 1.02 \times 10^{-6}$ sec) (Supplemental Material, Figure S4) [44], unreasonably long for the material's nature. This $\tau_0$ falls in the range expected for superspin glasses (~10$^{-6}$ sec) entailing magnetic nanoparticles [58,59] with dipole-dipole interactions instead of the shorter spin-flip times (10$^{-9}$-10$^{-13}$ sec) [42] anticipated for bulk systems with atomic $Mn^{3+}/Mn^{4+}$ magnetic moments.

The inadequacy of both Arrhenius and Vogel-Fulcher laws to describe the collective spin-freezing close to $T_f$, led us to infer that even stronger magnetic interactions may govern the spin-dynamics in the Na-birnessite compound and are the likely reason for the observed deviations from the phenomenology described above. The dynamics of a spin-glass system with strong underlying magnetic interactions is in this case adequately described by the dynamic scaling theory.[60] This approach, based on the standard theory for the dynamical slowing down of spin dynamics near the freezing point $T_f$, considers a divergence of the relaxation time ($\tau = 1/\omega$) at a finite temperature ($T_f$), which can be described by the following critical power-law:

$$\tau = \tau_o \left(\frac{T_f}{T_m - T_f}\right)^{zv} \qquad (4)$$

where $\tau_o$ is the characteristic attempt time that corresponds to the time needed for a single spin to undergo a transition from the paramagnetic to the frozen spin-glass state, $T_m$ is the frequency dependent $\chi'(T)$ peak-temperature position, and $T_f$ is the static spin glass transition temperature (equivalent to the freezing temperature when f→0), while the critical exponent, $zv$, has acceptable values between 4 and 12 for typical strongly interacting spin glasses [42].

The power-law, scaling analysis of the frequency dispersion of the $\chi'(T)$ maximum is presented as a linear fit [61] in the inset of Figure 10a. The value of $T_f$ has been adjusted manually in order to obtain the best linear fit in the log($\tau$) versus log[$T_f/T_m$-$T_f$] plot, while the $zv$ and $\tau_o$ were derived from the least-square fitting (Supplemental Material, S4) [44]. The best fit for the Na-birnessite yielded the following quantities: $T_f$ = 29.64 K, $\tau_o$ = 4.38×10$^{-13}$ (±0.49×10$^{-13}$) sec and $zv$ = 5.0(1). Previous investigations have suggested that some variation of these dynamic critical scaling parameters may infer materials of differing spin dimensionality [62-64]. For example, short-range insulating spin glasses, such as $Fe_{0.5}Mn_{0.5}TiO_3$ having Ising character and the Heisenberg-like (isotropic) $CdCr_{1.7}In_{0.3}S_4$, present a $zv$ = 10.5 and $zv$ = 7, respectively, whereas the characteristic attempt times of spin-flip may be much larger than $10^{-12}$ sec for the former and smaller than that for the latter. The aforementioned analysis of the temperature-shift of the AC susceptibility cusp with the frequency ($\omega$), together with the evidence for irreversibility (bifurcation) in the DC susceptibility, corroborate that Na-birnessite can be thought of as a magnetic system that resembles the cooperative behavior of other archetypical spin glasses (cf. CuMn, $zv$ = 5.5 and $\tau_o \sim 10^{-12}$ sec) [42,62,65].

### F. Field-Dependence of Magnetic Susceptibility

In order to further investigate the low temperature transition and bearing in mind that in a spin-glass the susceptibility cusp is very sensitive to an external applied magnetic field, $\chi(T)$ was studied under such a stimulus. Edwards and Anderson [66] predicted that the susceptibility field-dependent transition cusps become smeared out even under low magnetic fields. Indeed, this phenomenon is confirmed for the Na-birnessite. Figure 11 presents the ZFC-FC χ(T) data measured at magnetic fields up to 100 mT. The sharpness of the transition is smeared out when the field is increased progressively (~20 mT) until it becomes almost flat for fields of ~100 mT. Similar behavior has been observed in a number of magnetic systems that undergo a spin-glass freezing, such as the perovskites $La_{0.5}Sr_{0.5}CoO_3$[39] and $Eu_{0.58}S_{0.42}MnO_3$[67], as well as in chemically diverse intermetallic compounds, such as $PrRuSi_3$ [68]. In addition, the transition temperature, $T_f$, is shifted to lower values as the external magnetic field increases. A possible phase diagram for the crossover between paramagnetic and glassy states can then be roughly sketched (Figure 12). This DC field-dependence of $T_f(H)$ [39,50,69], has been assessed on the basis of the mean-free model of Sherrington and Kirkpatrick [70]:

$$\frac{T_f(0)-T_f(H)}{T_f(0)} \propto H^\delta \quad (5)$$

It broadly represents a critical line that defines the onset of irreversibility when entering the spin glass state in the H-T phase space. In a Heisenberg (isotropic) spin system, the transition occurs along the Gabay-Toulouse [71] (GT) line for $\delta = 2$ [72], while for an Ising type, the transition occurs along the de Almeida-Thouless [73] (AT) line for values of $\delta = 2/3$. From the least-square fitting (inset, Figure 12), Na-birnessite appears to follow the AT-like line of $H^{2/3}$, instead of GT line (Supplemental Material, S6) [44] for isotropic spins, a behavior that might have been inferred by the analysis of the dynamical scaling exponent (*zv*) and the Mydosh parameter. This likely discrepancy may be reconciled by claiming the role of weak, random anisotropy (e.g. Dzyaloshinsky-Moriya, single-ion etc) that is usually present in real systems [74,75]. Mean-field calculations have predicted that when anisotropic interactions are present, an otherwise Heisenberg spin-glass exhibits an Ising-like behavior in the low-field limit [76,77]. This has been verified experimentally in spin glasses of diverse chemical nature, ranging from the archetype metallic Ruderman-Kittel-Kasuya-Yosida (RKKY) CuMn [78] system to short-range interacting, insulating $CdCr_{1.7}In_{0.3}S_4$ [79] and semiconducting $Cd_{0.62}Mn_{0.38}Te$ [80] compounds. The present experimental data in Na-birnessite could suggest that anisotropic interactions should be taken into account, but on the other hand, can also imply that such conventional experiments may not be sufficiently sensitive [75] to the freezing of transverse spin degrees of freedom that are identifying features of Heisenberg spin glasses, consistent with a transition line of the GT type.

Nevertheless, further experimental evidence from the critical behaviour in the field-temperature phase diagram and the associated critical dynamics (e.g. isothermal time decay of AC susceptibility), may be required in order to verify the dimensionality of the spin interactions in the system under study.

**G. Interactions in Related Manganese Dioxides**

An interesting reference system, owing to its chemical affinity, is the hexagonal $Na_{0.7}MnO_2$ ($P6_3/mmc$; a= 2.811 Å, c= 11.118 Å) layered derivative that has been identified to undergo a spin-glass transition at $T_f \leq 39$ K [43] due to random distribution of the $Mn^{3+}/Mn^{4+}$ cations in a frustrated triangular lattice topology. Consequently, glassiness develops and reflects on the DC susceptibility, with common features between the two materials pertaining to the ZFC-FC divergence in $\chi(T)$, the elimination of the ZFC-FC $\chi(T)$ maximum upon the application of moderate external magnetic fields (H< 500 mT), and the comparable Mydosh parameter, which was calculated as K= 0.004 for $Na_{0.7}MnO_2$. Interestingly, the bifurcation of the ZFC/FC susceptibilities below $T_f$ indicates weakly coupled ferromagnetic species for both, with a comparable coercive field of about 0.1 T in the 5-10 K region (Figure S5 and Figure 4 of ref. [43]). To rationalize this, the different degree of $Mn^{3+}:Mn^{4+}$ disproportionation in the two systems needs to be considered, namely, 0.86:0.14 for $Na_{0.7}MnO_2$ versus 0.51:0.49 for Na-birnessite (Supplemental Material, S2) [44]. Although a rigorous evaluation of the relative magnitude of interactions between any pair of Mn-cations is difficult given the large number of individual microscopic processes involved in the magnetic exchange pathways, one expects that the increased abundance of $Mn^{4+}$ states in Na-birnessite will enhance ferromagnetic interactions (as $Mn^{4+}$-O-$Mn^{4+}$ 90-degree superexchange is ferromagnetic) [81], in analogy to the closely related layered $Na_{5/8}MnO_2$ [23]. Consequently, with the weight of the interactions likely shifted to ferromagnetic-like (against antiferromagnetic) correlations, the reduction in the Curie-Weiss temperature for Na-birnessite is also anticipated (i.e. -64 K vs. -411 K for $Na_{0.7}MnO_2$). The material offers a test-bed to probe subtle details of competing processes in multivalent Mn-systems, thus going beyond cooperative magnetism and possibly into emerging rechargeable battery technologies. In view of the latter, electron-driven (Jahn-Teller) lattice transformations met in $Na_xMnO_2$ derivatives deserve to be studied as a reason to understand how to avoid impairing [23,82,83] the electrochemical properties of candidate layered oxides for sodium-ion rechargeable cathodes.

*On the Origin of the Spin Glass State in Na-birnessite.*

In chemically related dioxides, such as $Na_{0.7}MnO_2$ and $\alpha$-$K_{0.087}MnO_2$, with frustrated triangular Mn topology, alkali meal off-stoichiometry in the interlayer sites imposes disorder in the manganese sub-lattice by random generation of $Mn^{3+}$ and $Mn^{4+}$ states (site disorder) [43,48]. This, promotes bond disorder (i.e. the sign of the exchange interaction between nearest-neighbors may flip at random from AFM to FM and vice-versa), and therefore raises frustration [84]. In view of the latter, it is useful to highlight some of the observed microscopic characteristics of the Na-birnessite structure. The afore-mentioned avenues for site disorder (and frustration) in the compound under the present study would be enhanced due to possible vacancies in the $MnO_6$ layers. For example, the Mn vacancies, whether attributed to the $Mn^{3+}$ or $Mn^{4+}$ cations, would result in the disruption of the long-range periodicity of the magnetic interactions. Imperfections of the crystal structure and specifically, stacking-faults were indeed revealed by the HRTEM experiments (Figure 2c). Planar defects of this type have a significant impact as they alter the exchange pathways, the relative number and strength of exchange interactions as well as the type of the magnetic coupling (AFM or FM). Under these conditions the development of various competing exchange interactions and their interplay would give rise to frustration. An empirical way to estimate the degree of frustration is to calculate the ratio f= -$\theta_{cw}/T_N$, where $T_N$ is a temperature at which the system would order cooperatively. The values of f>1 imply that the material is magnetically frustrated [85]. Regarding the compound under study, f = 2.21 (with $\theta_{cw}$ = -64.28 K and $T_N$ = 29 K), a value in accord with the previous arguments. Although the triangular in-plane lattice topology triggers magnetic frustration *per se*, the simultaneous presence of $Mn^{3+}/Mn^{4+}$ cations that are randomly distributed within the $MnO_6$ layers is an additional avenue generating competing exchange interactions, in turn promoting spin frustration and collective spin-glass freezing in the hydrated variant of the $\alpha$-$NaMnO_2$ antiferromagnet.

## V. CONCLUSIONS

In summary, we have shown a paradigm where facile, hydration preparative routes to layered sodium manganese dioxides have driven significant alteration of the crystal structure, leading to fundamental changes in the electronic character and the magnetic state, therefore opening prospects for enhanced functional behavior. Specifically, the insertion of $H_2O$ molecules in the interlayer spacing of the two-dimensional frustrated antiferromagnet α-$NaMnO_2$ leads to the formation of the $Na_{0.36}MnO_2 \cdot 0.2H_2O$ derivative with enhanced $MnO_6$-layer separation and mixed-valency that offers favorable redox attributes for technologically useful, rechargeable battery cathode materials. From fundamental point of view, the intercalation of water in α-$NaMnO_2$ has a dramatic impact on the antiferromagnetic ground state of the parent compound ($T_N$= 45 K). Na-birnessite is shown to enter a cooperative spin-frozen state, at $T_f$= 29 K, where the critical dynamics of strongly interacting spins is governed by the divergence of their relaxation times. This is in agreement with the power-law, scaling theory that dictates the magnetic dynamics before the collective spin-glass freezing. The variation in the magnetic response, imposed by the structural transformation and the chemically driven in-plane defect environment due to crystal $H_2O$, highlight the critical impact of topology and site disorder on frustrated magnetism.

## ACKNOWLEDGMENTS

This research has in part been co-financed by the European Union and Greek national funds through the Research Funding Program Heracleitus II (grand number 349309 WP1.56). Part of this work has been performed in the framework of the PROENYL research project, Action KRIPIS, project No MIS-448305 (2013SE01380034) that was funded by the General Secretariat for Research and Technology, Ministry of Education, Greece and the European Regional Development Fund (Sectoral Operational Programme: Competitiveness and Entrepreneurship, NSRF 2007-2013)/ European Commission. We acknowledge the support of the National Institute of Standards and Technology, U. S. Department of Commerce, in providing the neutron research facilities used in this work. AT acknowledges financial support from the Federal Ministry for

<="">Education and Research through the Sofja Kovalevskaya Award of Alexander von Humboldt Foundation.</>

**FIGURES**

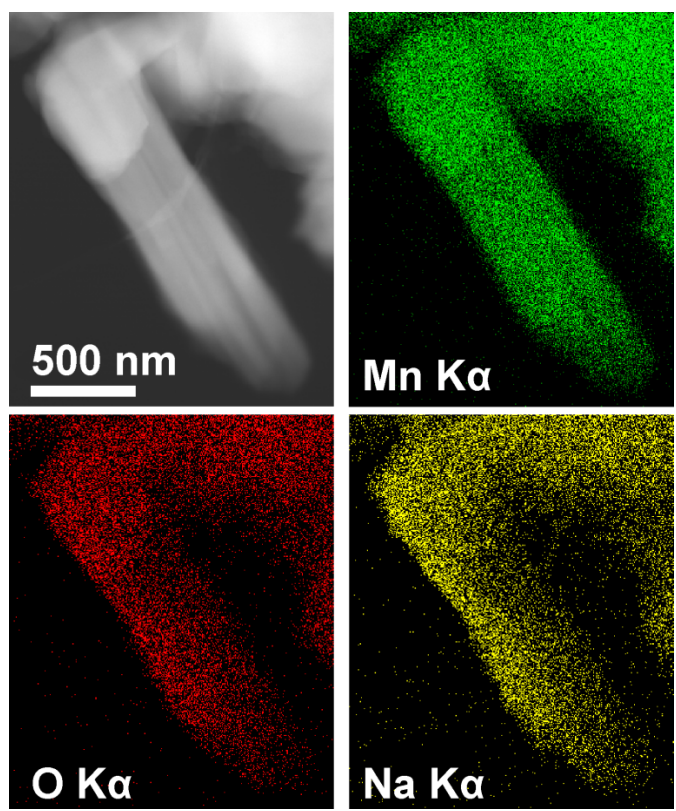

**FIG. 1.** HAADF-STEM image of a typical Na-birnessite crystal and corresponding EDS maps showing the homogeneous distribution of Mn, Na and O over it.

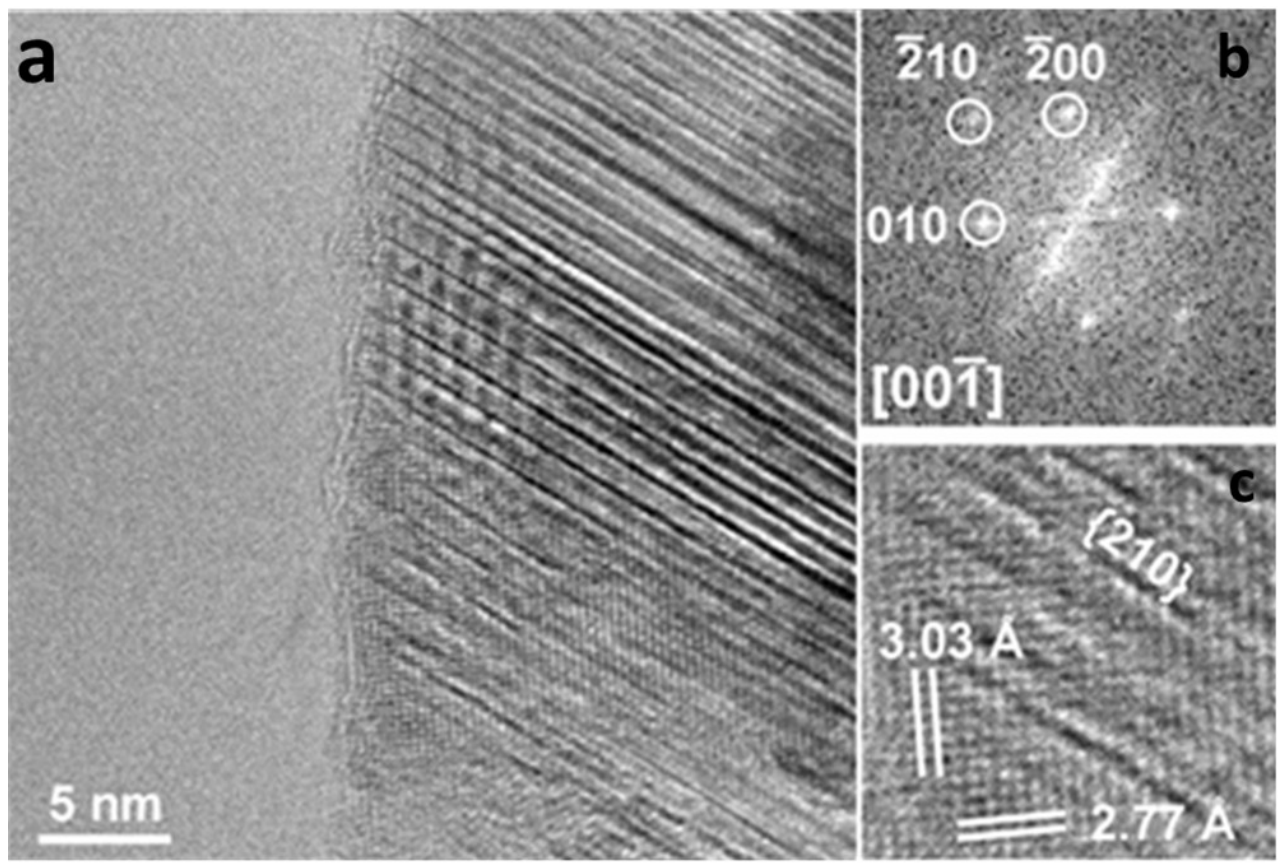

**FIG. 2.** (a) HRTEM image of a typical crystallite in the Na-birnessite sample, suspended in vacuum and (b) fast Fourier transform (FFT) of the area in (c). The observed diffraction spot-pattern matches the [122]-oriented triclinic $Na_{0.3}MnO_2 \cdot 0.93H_2O$ lattice, with 6% cell expansion.

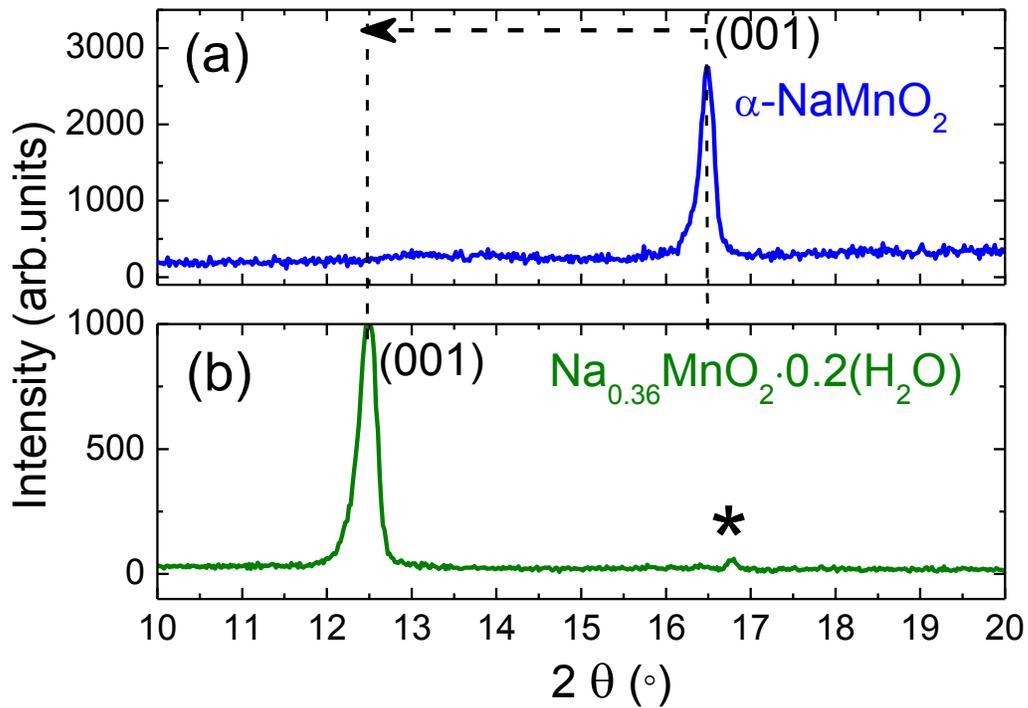

**FIG. 3.** Low-angle part of the XRPD patterns for α-NaMnO$_2$ (a) and the Na-birnessite (b), showing the characteristic shift of the 001 Bragg reflection upon hydration; (*), non-converted parent α-NaMnO$_2$.

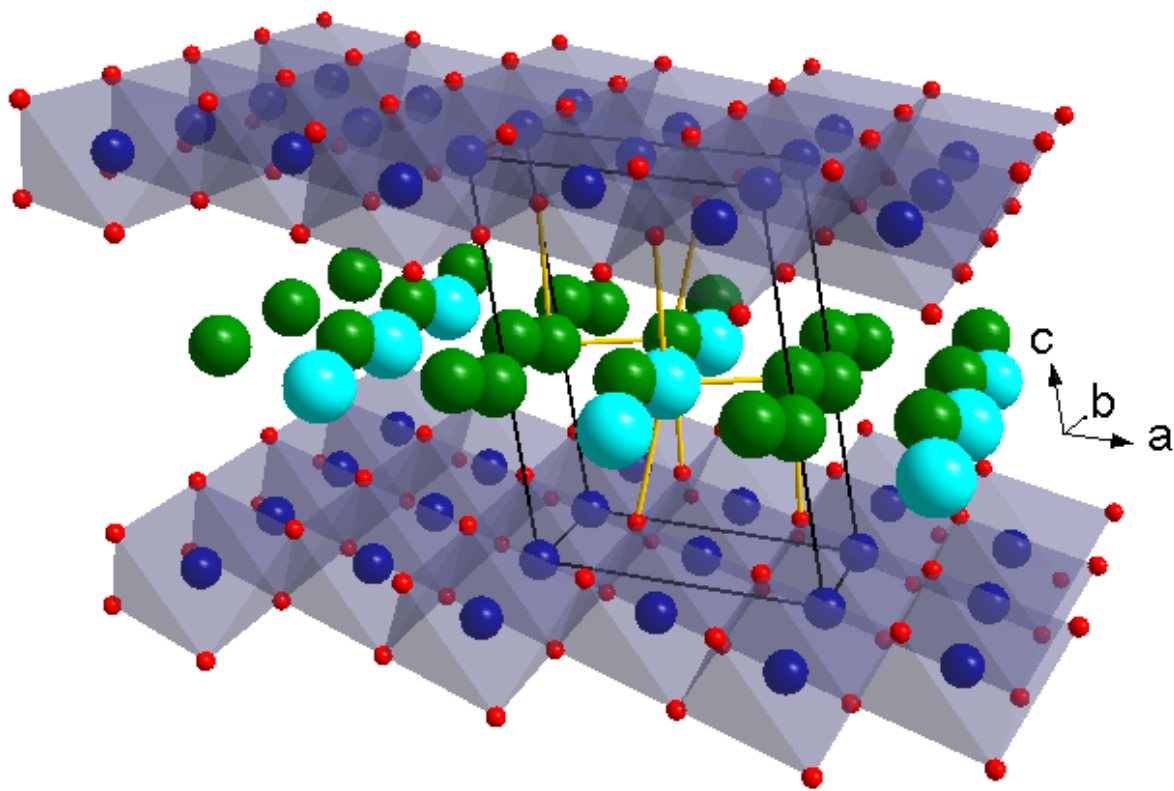

**FIG. 4.** Schematic representation of the crystal structure of the Na-birnessite: the manganese, oxygen, sodium atoms are represented with the blue, red and green spheres; the $H_2O$ molecules in the interlayer space are shown with the turquoise spheres; cell-edges are represented with the black lines. For ease of visualization, Na atoms and $H_2O$ molecules, occupying identical positions in the reported structure, are displayed in alternating positions in the model (ICSD 262208).

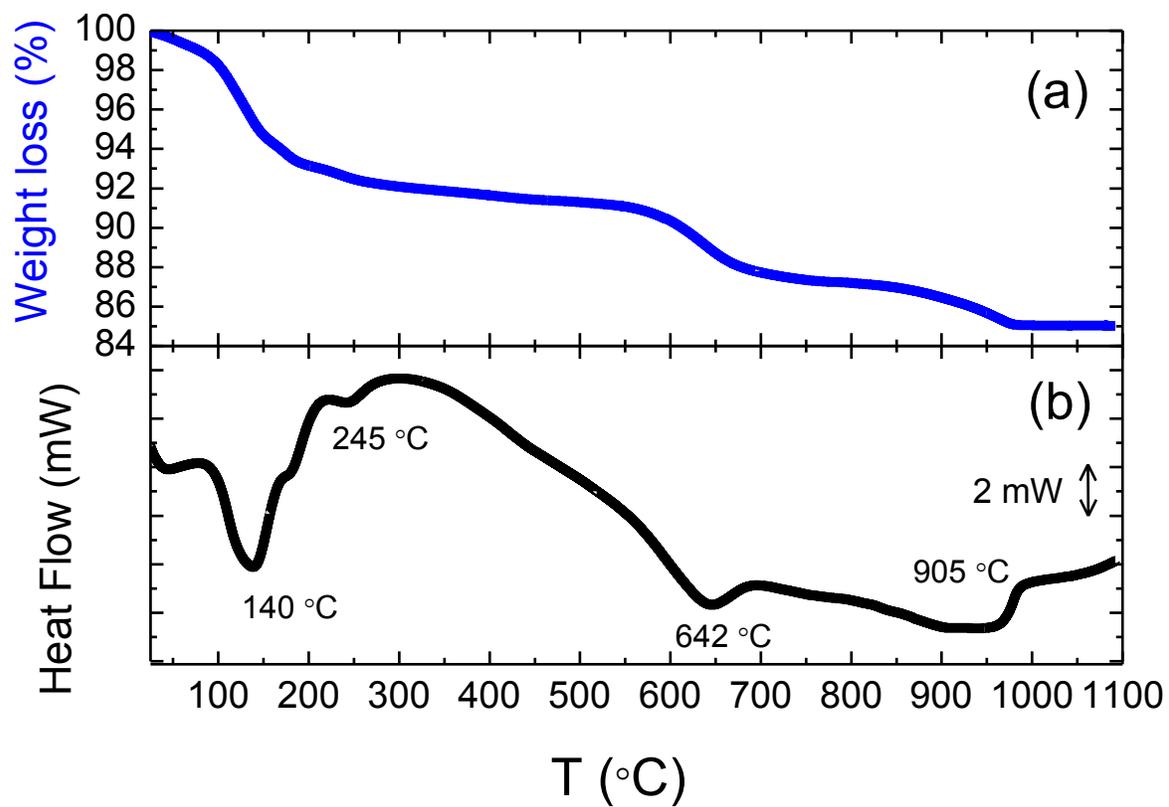

**FIG. 5.** (a) Thermogravimetric (TGA) and (b) heat-flow (DTA) measurements for the Na-birnessite up to 1100 °C.

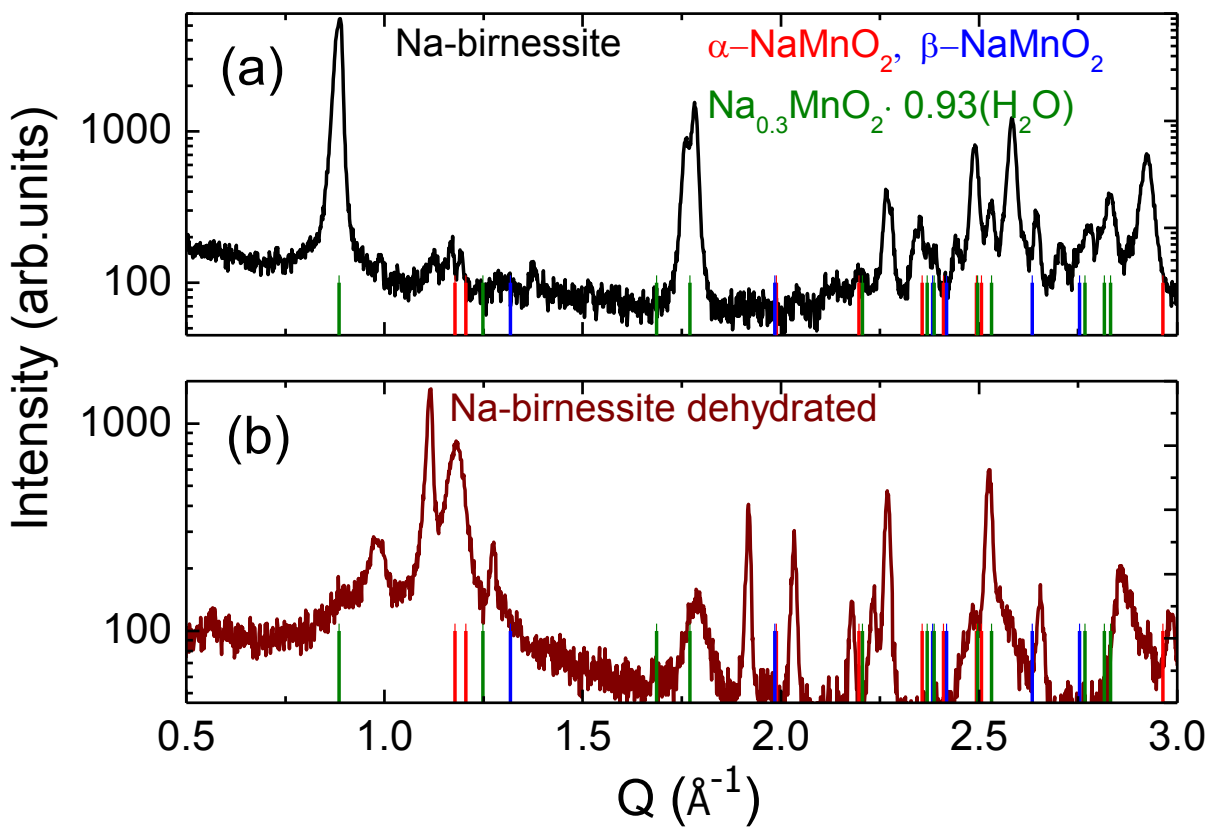

**FIG. 6.** Comparison of the XRPD patterns of the $Na_{0.3}MnO_2 \cdot 0.2H_2O$ birnessite's polycrystalline powder (a) and the dehydrated compound (b) taken after the TGA experiment performed up to 1100° C. Indexing of the reflections has been done on the basis of the α-$NaMnO_2$ (ICSD 16270), β-$NaMnO_2$ (ICSD 16271) and the $Na_{0.3}MnO_2 \cdot 0.93H_2O$ (ICSD 262208) phases, whose Bragg reflections are indicated with red, blue and green tick marks, respectively. Bragg peaks attributed to the $Mn_3O_4$ (ICSD 68174, ICSD 1514104) are left without tick marks for ease of comparison.

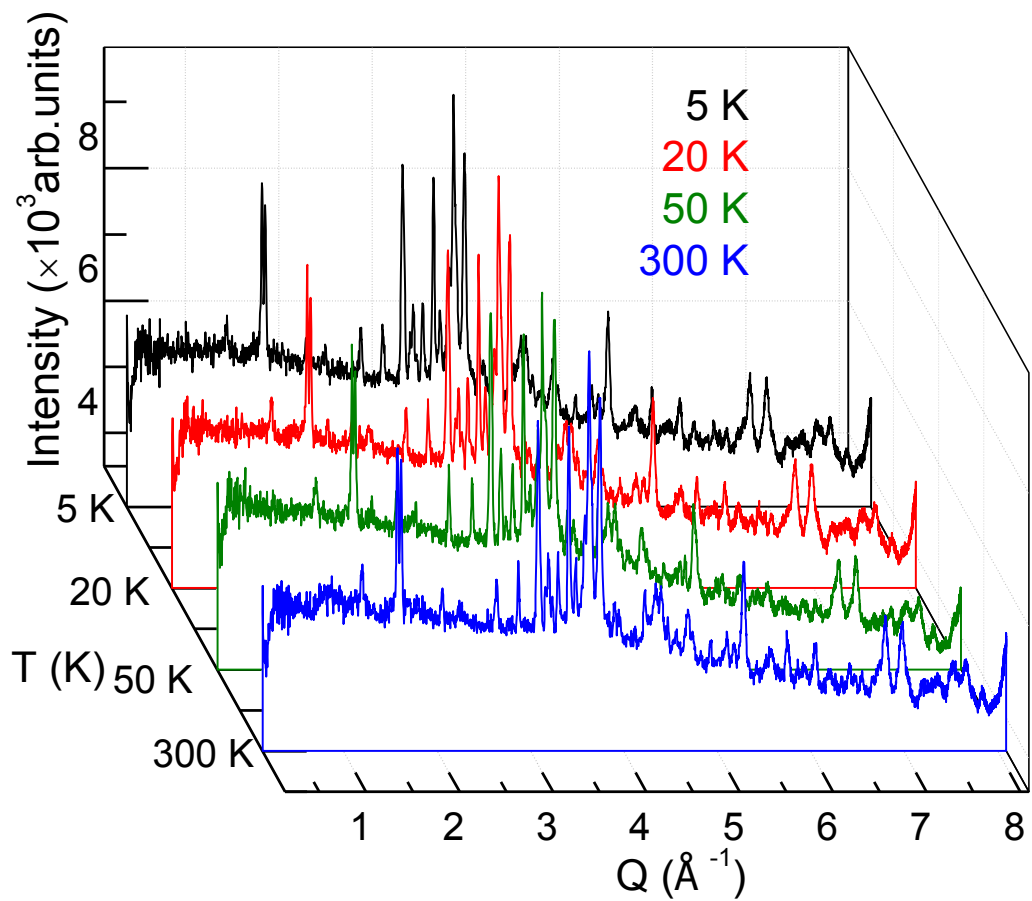

**FIG. 7.** A series of neutron powder diffraction patterns for Na-birnessite obtained at 5, 20, 50 and 300 K. The absence of reflections of magnetic origin in the low-temperature region, indicates that no long-range magnetic order develops below 29 K.

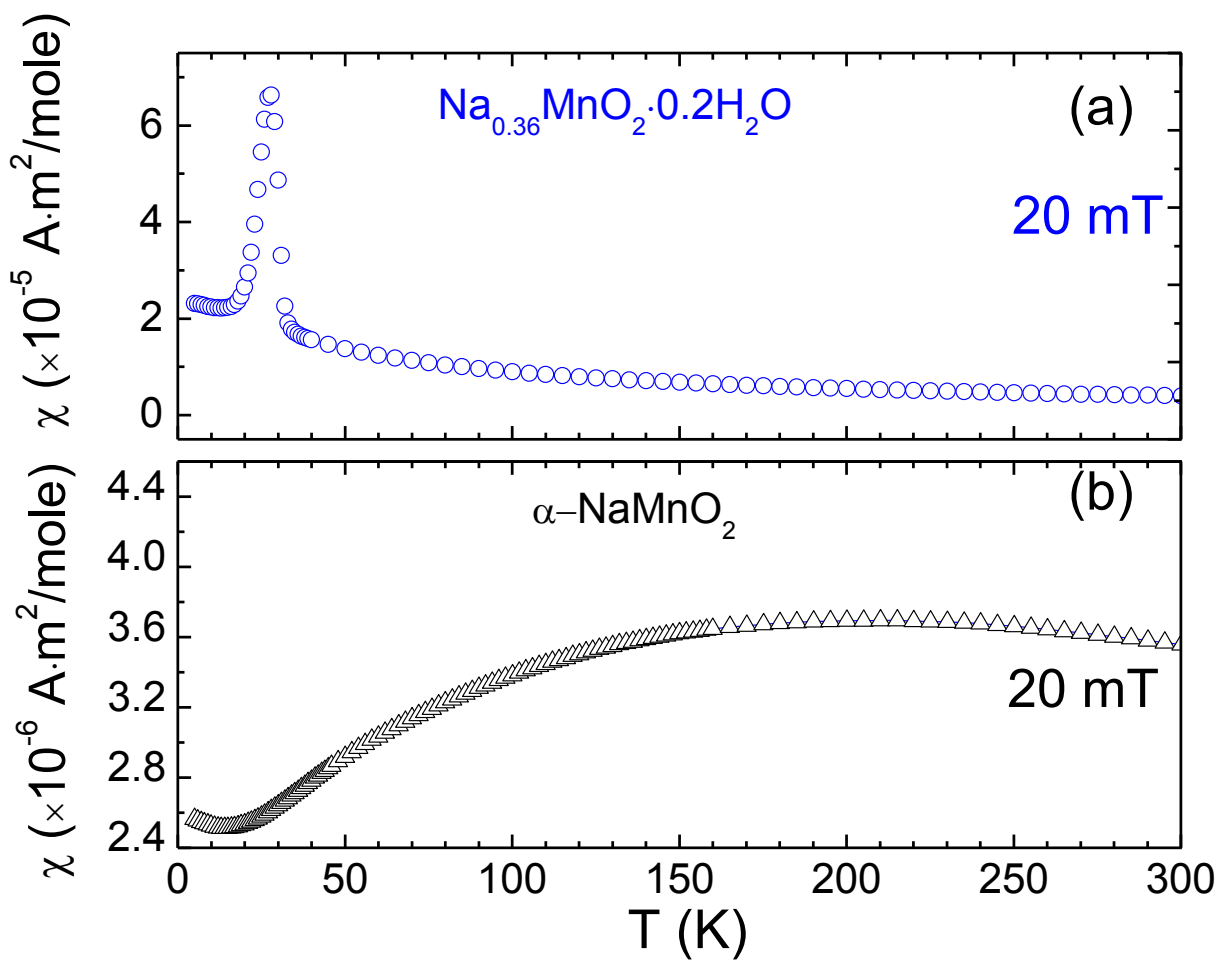

**FIG. 8.** The temperature evolution of the zero-field cooled (ZFC) DC magnetic susceptibility for (a) α-NaMnO$_2$ and (b) Na-birnessite, under an applied field of 20 mT.

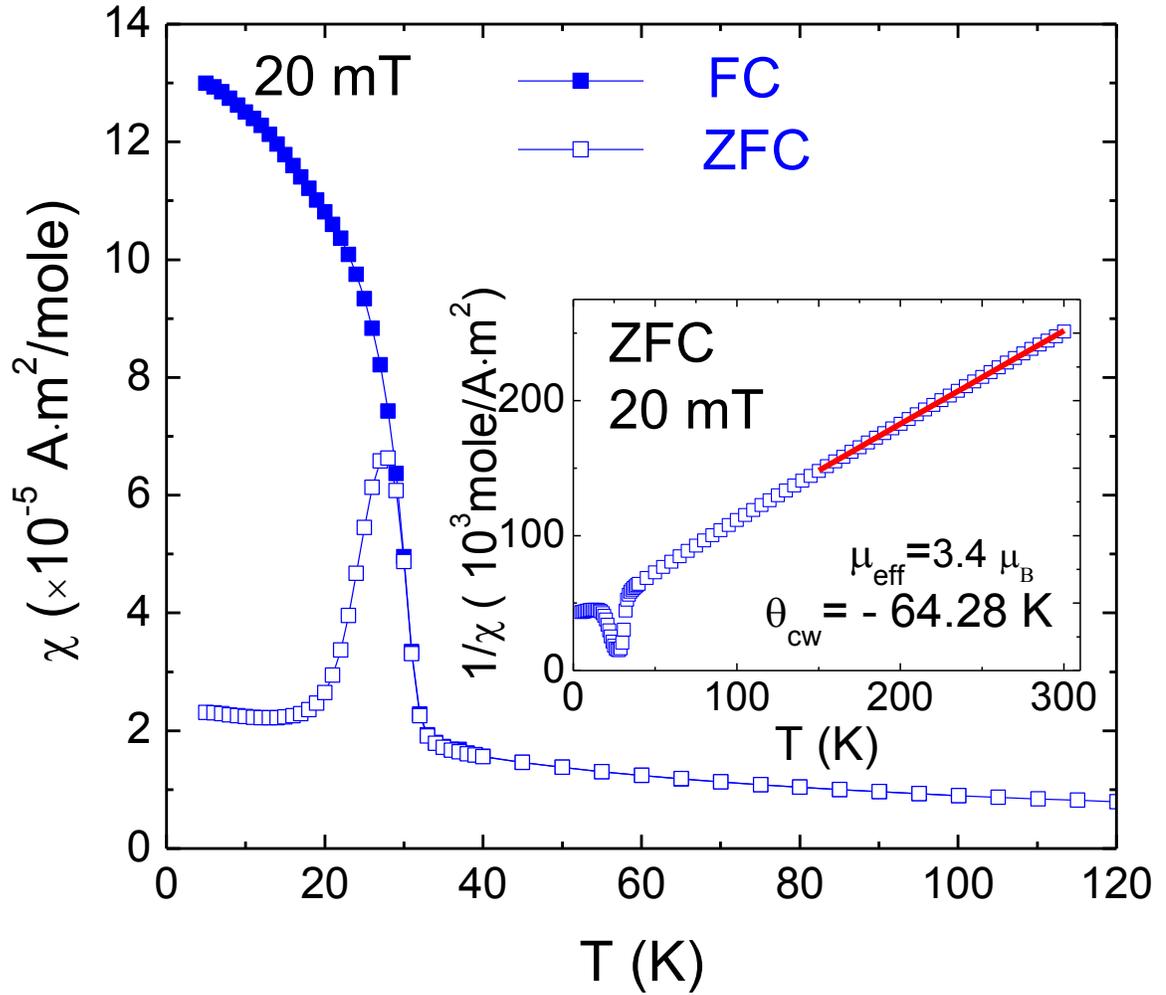

**FIG. 9.** Low-temperature evolution of the ZFC-FC DC susceptibilities for the Na-birnessite. Note: Characteristic sharp peak at $T_f = 29$ K in the ZFC protocol and bifurcation between the ZFC and FC curves below the $T_f$. Inset: Curie-Weiss fit (red continuous line) of the reciprocal susceptibility (blue data points).

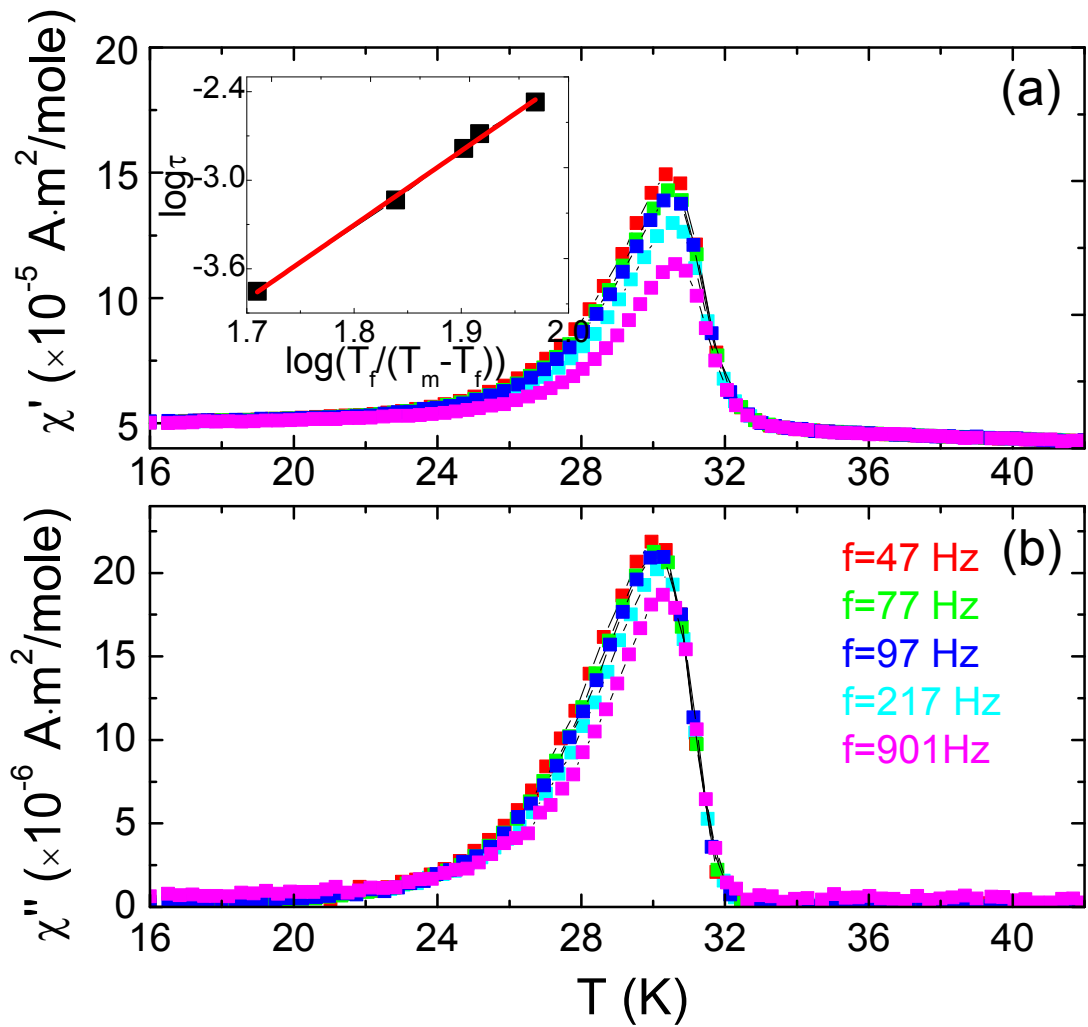

**FIG. 10.** Real ($\chi'$) and imaginary ($\chi''$) parts of the AC susceptibility versus temperature at five different frequencies, under an AC drive field 0.3 mT. Inset: Power-law, scaling analysis of the frequency dispersion of the real part $\chi'$ (T) peak temperature position

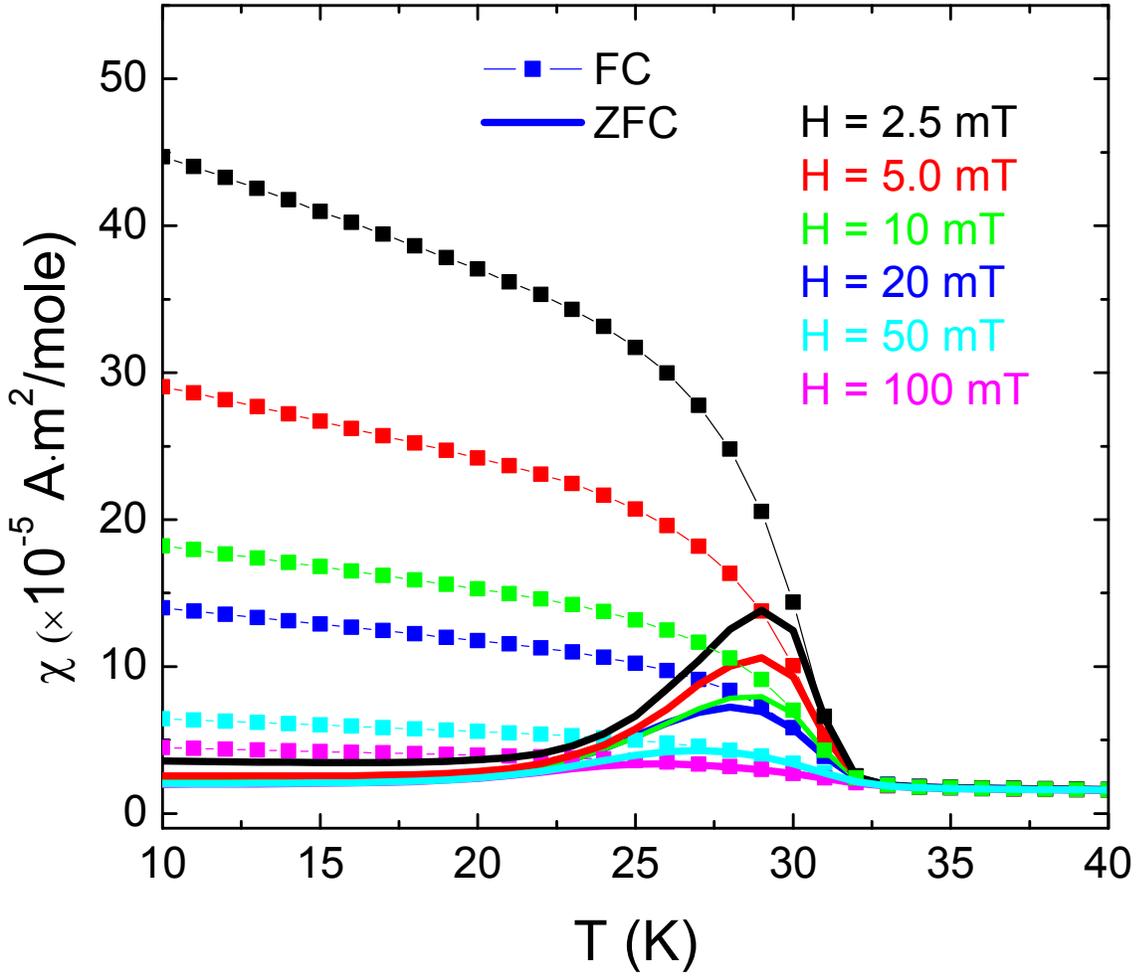

**FIG. 11.** Field-dependence of the ZFC-FC DC magnetic susceptibility of the Na-birnessite measured under magnetic fields ranging from 2.5 - 100 mT. The 'cusp' at 29 K is eliminated as the external magnetic field is increased.

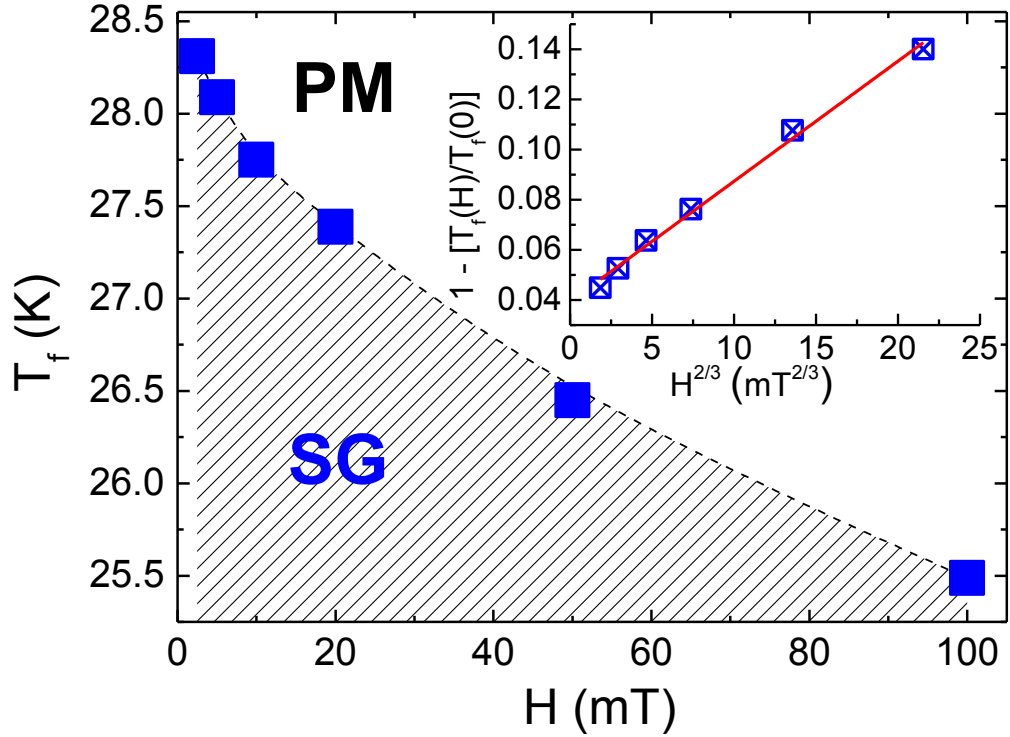

**FIG. 12.** The magnetic field dependence of the freezing temperature ($T_f$) in the Na-birnessite depicts a relevant phase-diagram for the transition between the PM (paramagnetic) and SG (spin glass) states. Inset: Least square fit of the reduced temperature in the de Almeida-Thouless mean-free model (equation 5).

.

# Supplemental Material

## Hydration induced spin glass state in a frustrated Na-Mn-O triangular lattice

Ioanna Bakaimi,[1,2] Rosaria Brescia,[3] Craig M. Brown,[4,5] Alexander A. Tsirlin,[6] Mark A. Green,[7] and Alexandros Lappas[1,*]

[1] *Institute of Electronic Structure and Laser, Foundation for Research and Technology – Hellas, Vassilika Vouton, 71110 Heraklion, Greece*

[2] *Department of Physics, University of Crete, Voutes, 71003 Heraklion, Greece*

[3] *Nanochemistry Department, Istituto Italiano di Tecnologia, Via Morego 30, 16163 Genova, Italy*

[4] *NIST Center for Neutron Research, 100 Bureau Drive, Gaithersburg, MD 20899-8562, USA*

[5] *Department of Chemical and Biomolecular Engineering, University of Delaware, Newark, DE 19716, USA*

[6] *Experimental Physics VI, Center for Electronic Correlations and Magnetism, Institute of Physics, University of Augsburg, 86135 Augsburg, Germany*

[7] *School of Physical Sciences, University of Kent Canterbury, Kent CT2 7NH, UK*

**Table of Contents:**

## S1. Supplementary images from electron microscopy studies

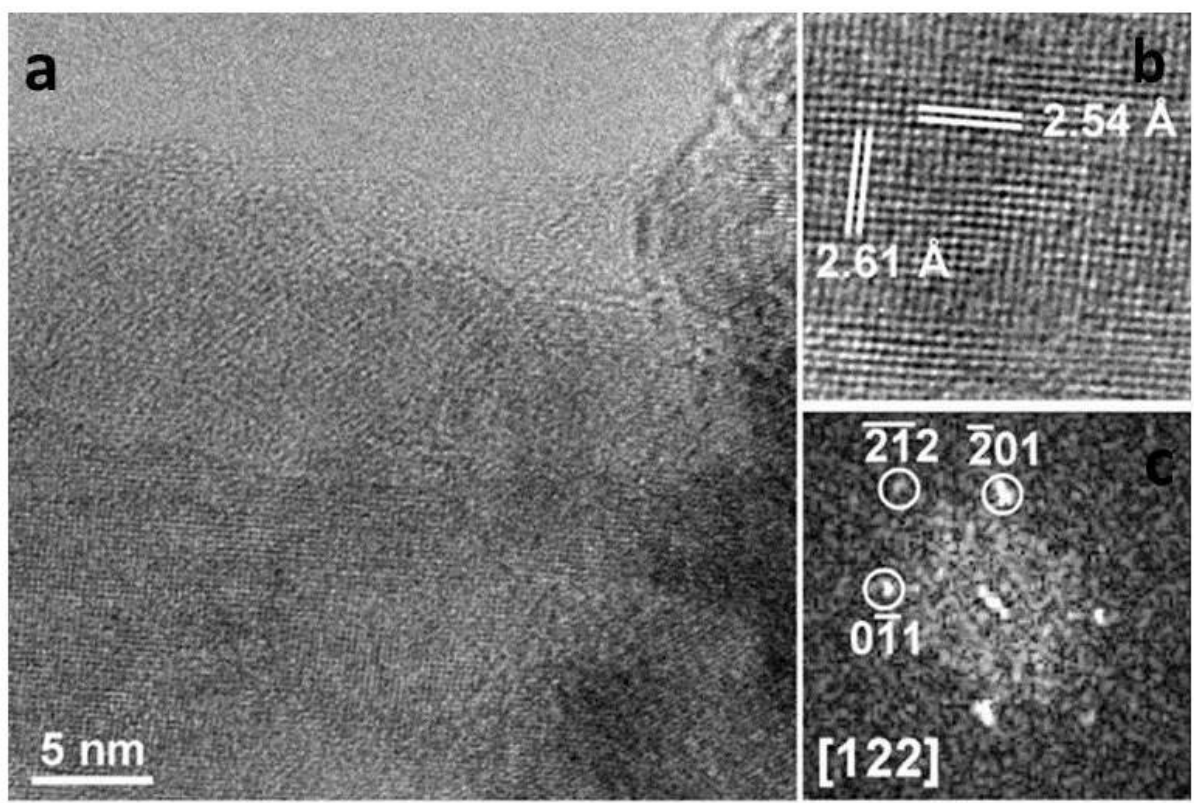

**FIG. S1.** HRTEM image showing a typical crystallite of the $Na_{0.36}MnO_2 \cdot 0.2H_2O$ compound. The observed diffraction spot-pattern matches with the [122]-oriented triclinic $Na_{0.3}MnO_2 \cdot 0.93H_2O$ [1], with 6% cell expansion.

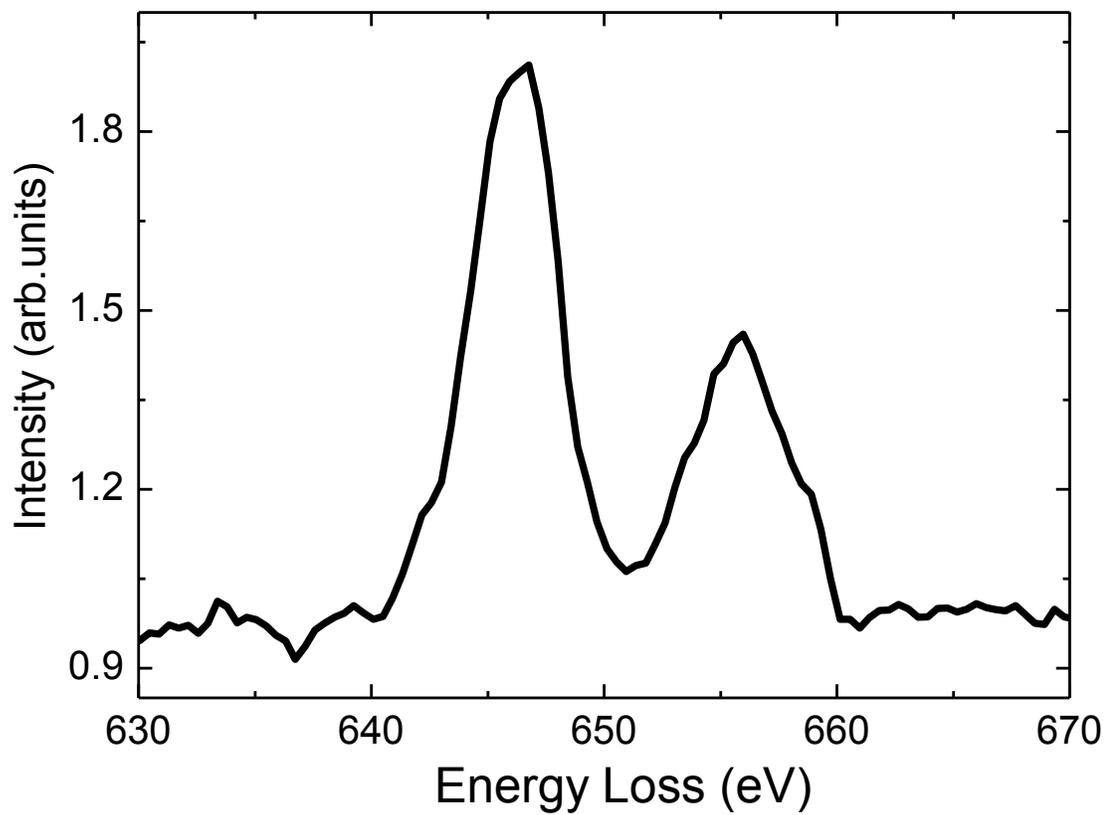

**FIG. S2**. Region of the Mn $L_{23}$ core-loss EELS spectrum analyzed for the determination of the oxidation state of Mn in the $Na_{0.36}MnO_2 \cdot 0.2\,H_2O$ compound.

## S2. Estimation of the ratio $Mn^{3+}:Mn^{4+}$

From the Curie-Weiss fit of the χ(T) and assuming an average spin-only contribution from $Mn^{4+}$ and $Mn^{3+}$, an effective moment of $\mu_{eff}$= 3.401 $\mu_B$ has been estimated for the Na-birnessite. If $Mn^{3+}$ was in a high-spin state (4.9 $\mu_B$) then the average effective moment could have been even larger than 3.87 $\mu_B$ expected for the spin-only configuration of $Mn^{4+}$. Therefore, it can be assumed that the $Mn^{3+}$ is likely contributing to the average effective moment with its low-spin state. In order to estimate the relative ratio of $Mn^{3+}/Mn^{4+}$ species we take into account the following equations:

$$\chi = x\,\chi_{Mn^{3+}} + y\,\chi_{Mn^{4+}} \qquad (1)$$

$$\chi = \frac{C}{T - \theta_{cw}}$$

$$C = \frac{N\,\mu_{eff}^2}{3k_B}$$

$$\mu_{eff}^2 = x\,\mu^2_{Mn^{3+}} + y\,\mu^2_{Mn^{4+}} \qquad (2)$$

where $\mu_{eff}(Mn^{3+})$= 2.83 $\mu_B$, $\mu_{eff}(Mn^{4+})$ = 3.87 $\mu_B$ and x= 1-y  (3)

Using relations (2) and (3) we get: x= 0.506 and y= 0.494, thus $\frac{Mn^{3+}}{Mn^{4+}} \approx 1.025$  (4)

## S3. AC susceptibility data analyses: Arrhenius and Vogel-Fulcher laws

The frequency dependence of the χ'(T) maximum of the data shown in Figure 10 have been also analyzed by the phenomenological Arrhenius (Figure S3) and the Vogel-Fulcher (Figure S4) laws. The analysis did not yield parameters with physically acceptable values, with a critical power lad providing better description (see main manuscript).

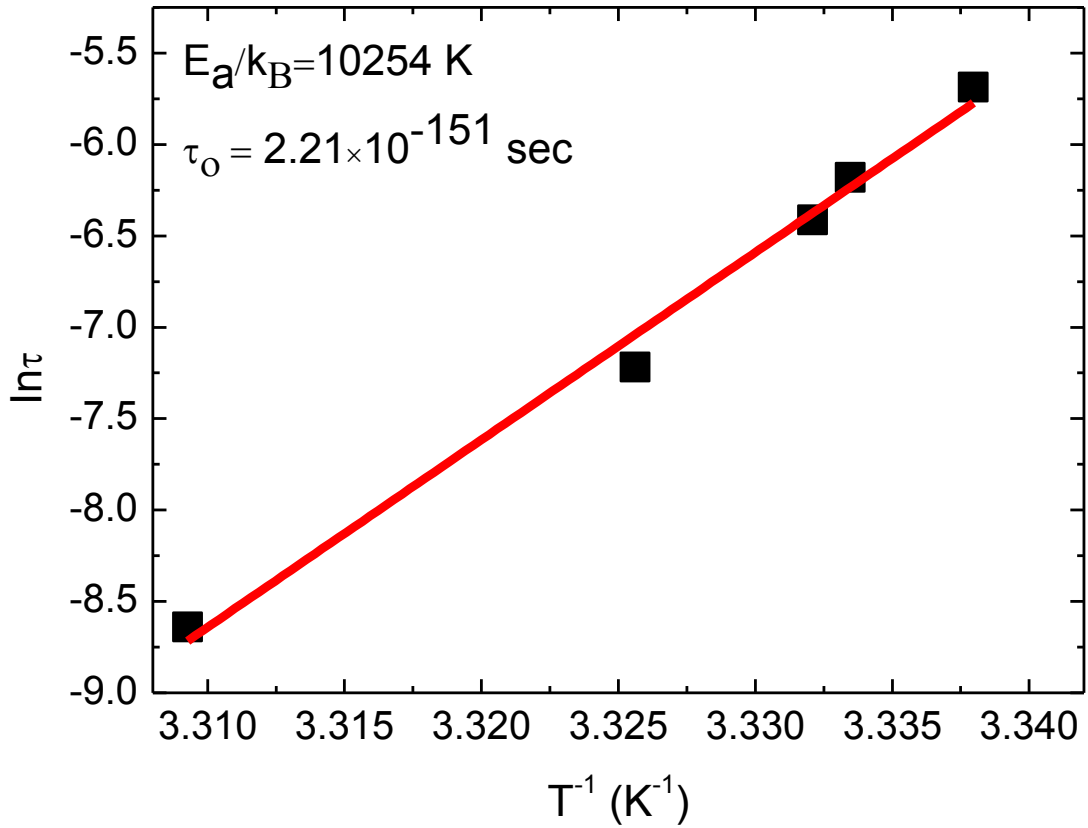

**FIG. S3.** The Arrhenius law description of the frequency dispersion of the χ'(T) peak-temperature position (data of Figure 10). The fit yielded physically unreasonable values for the activation energy, $E_a$, and the attempt time, $\tau_o$, excluding the possibility of superparamagnetism for the $Na_{0.36}MnO_2 \cdot 0.2H_2O$.

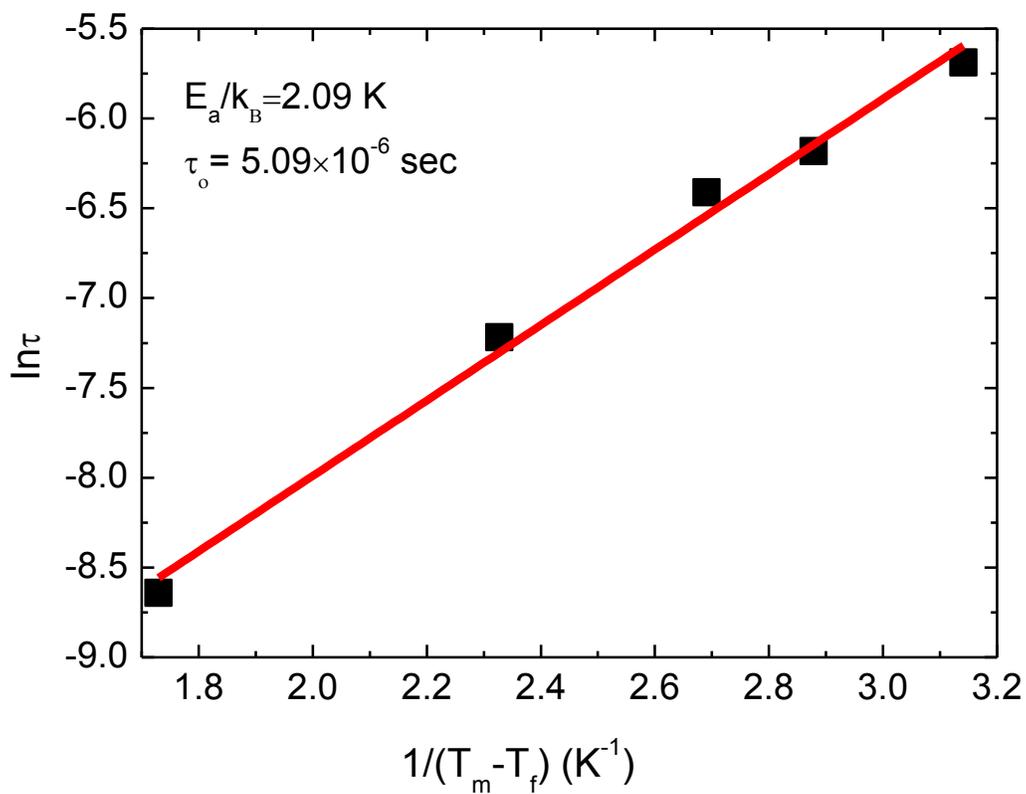

**FIG. S4** The Vogel-Fulcher law description of the frequency dispersion of the $\chi'(T)$ peak-temperature position (data of Figure 10). The unusually short resultant attempt time ($\tau_o$) rules out the possibility of intermediate strength interactions for the Na-birnessite.

## S4. AC susceptibility data analyses: Power-law

We discuss the methodology used for the estimation of the parameters $zv$, $\tau_o$ and $T_f$ when the power-law (equation 4) was employed. Here $T_m$ is the temperature of the susceptibility cusp that changes upon frequency change. In order to estimate the peak-temperature position at each different frequency all the curves of the real part of the magnetic susceptibility ($\chi'(T)$), shown in Figure 13, were fitted with a Lorentz function. $T_f$ refers to the spin-glass transition when f→0. Initially one notices that there are three parameters that should be defined: the $zv$ that corresponds to the slope of the linear fit, the attempt time $\tau_o$ and the temperature $T_f$. To reduce the complexity of a three parameter linear fit, we follow a previously suggested method that calculates two parameters [2]: the $zv$ and the $\tau_o$. The value of $T_f$ is adjusted manually in order to get the best linearity in the log ($\tau$) versus log[$T_f/(T_m-T_f)$] plot. The $zv$ and $\tau_o$ are calculated from the extracted parameters (slope and intercept) of the least square fit. The chosen value of $T_f$= 29.64 K resulted in the best linear fit (Adj.R-Square= 0.997), so the $zv$ and $\tau_o$ were extracted accordingly. As $T_f$ is chosen and not calculated, we assume no estimated standard error for this parameter.

**S5. Hysterisis loop of the Na-birnessite obtained at 5 K.**

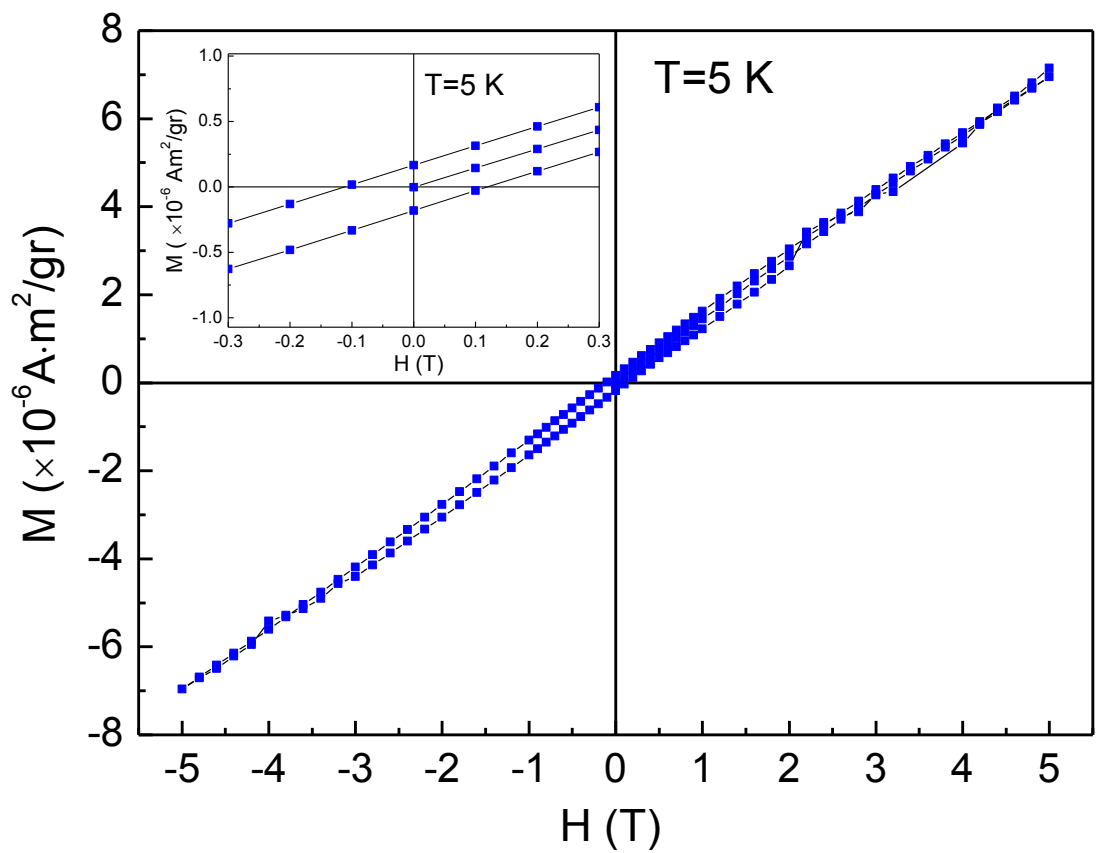

**FIG. S5.** Magnetization versus magnetic field of the $Na_{0.36}MnO_2 \cdot 0.2H_2O$ compound at 5 K. Inset: detail of the M(H) curve, suggests a coercive field $H_c \sim 0.1$ T.

## S6. Gabay-Toulouse analysis.

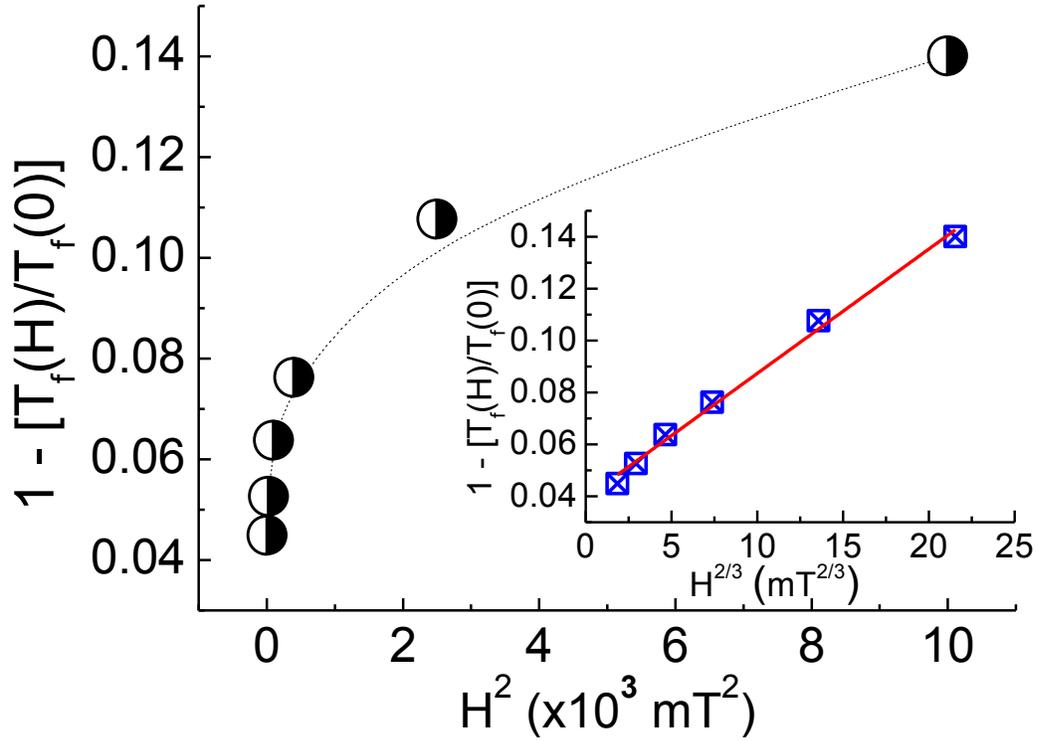

**Fig. S6.** Reduced freezing temperature (equation 5) on the basis of the Gabay-Toulouse (GT) model ($\delta = 2$) and the corresponding de Almeida-Thouless ($\delta = 2/3$) (inset).

The field dependence of the freezing temperature (data of Figure 11) according to the Gabay-Toulouse model deviates heavily from the linear behavior when $\delta = 2$ (equation 5), generally expected for a Heisenberg spin glass system.